\documentclass[fleqn,10pt]{wlscirep}
\usepackage[utf8]{inputenc}
\usepackage[T1]{fontenc}
\title{Robust Parallel Laser Driving of Quantum Dots for Multiplexing of Quantum Light Sources}

\author[1]{Ajan Ramachandran}
\author[1]{Grant R. Wilbur}
\author[1]{Reuble Mathew}
\author[1]{Allister Mason}
\author[2,+]{Sabine O'Neal}
\author[2,+]{Dennis G. Deppe}
\author[1,*]{Kimberley C. Hall}
\affil[1]{Department of Physics and Atmospheric Science,
Dalhousie University, Halifax, Nova Scotia B3H 4R2 Canada}
\affil[2]{The College of Optics and Photonics, University of Central Florida, Orlando, Florida 32816-2700, USA}

\affil[+]{Present Address of S. O'Neal: Imec, Kissimmee, Florida, 34744, USA. Present address of D. G. Deppe: SdPhotonics, Richardson, Texas 75081, USA.}
\affil[*]{Kimberley.Hall@dal.ca}
\begin{abstract}
Deterministic sources of quantum light (\textit{i.e.} single photons or pairs of entangled photons) are required for a whole host of applications in quantum technology, including quantum imaging, quantum cryptography and the long-distance transfer of quantum information in future quantum networks.  Semiconductor quantum dots are ideal candidates for solid-state quantum emitters as these artificial atoms have large dipole moments and a quantum confined energy level structure, enabling the realization of single photon sources with high repetition rates and high single photon purity. Quantum dots may also be triggered using a laser pulse for on-demand operation.  The naturally-occurring size variations in ensembles of quantum dots offers the potential to increase the bandwidth of quantum communication systems through wavelength-division multiplexing, but conventional laser triggering schemes based on Rabi rotations are ineffective when applied to inequivalent emitters.  Here we report the demonstration of the simultaneous triggering of $>$10 quantum dots using adiabatic rapid passage.  We show that high-fidelity quantum state inversion is possible in a system of quantum dots with a 15~meV range of optical transition energies using a single broadband, chirped laser pulse, laying the foundation for high-bandwidth, multiplexed quantum networks.      
\end{abstract}
\begin{document}

\flushbottom
\maketitle
% * <john.hammersley@gmail.com> 2015-02-09T12:07:31.197Z:
%
%  Click the title above to edit the author information and abstract
%
\thispagestyle{empty}

\section*{Introduction}

Owing to their strong optical transitions and the ability to tune their emission wavelengths, semiconductor quantum dots (QDs) have found applications in a variety of areas of optoelectronics, including lasers \cite{Park:1999}, light-emitting diodes \cite{Jang:2023} and solar cells \cite{Kamat:2013}.   These same properties also make QDs attractive for application to quantum emitters for secure quantum communications and quantum networks \cite{Senellart:2017}, with the performance of QD-based single photon sources continually improving over the past decade \cite{Senellart:2017,Arakawa:2020}.  Considerable progress has been made in the development of fabrication techniques, including photonic structures for efficient light extraction\cite{Haffouz:2018,Gerard:1998,Somaschi:2016,Takemoto:2007}, as well as effective excitation schemes for triggering the single photon emission \cite{Glassl:2013,Madsen:2014,Muller:2014,He:2019,Koong:2021,Bracht:2021,Wilbur:2022,Karli:2022}.  These advances have led to the demonstration of a single photon indistinguishability of 98.9\%, a high single-photon purity (with g$^{2}$(0)~=~7.5$\times$10$^{-5}$) and a collection efficiency of 83\% in recent years \cite{Somaschi:2016,Schweickert:2018,Laferriere:2022}.  In addition, telecom-compatible QD emitters have been developed that will aid in the long-distance transfer of quantum states using conventional optical fiber technology \cite{Haffouz:2018,Gamouras:2013,Nawrath:2019,Muller:2018}.   Such high-performance QD-based quantum light sources have enabled the recent demonstration of multi-photon graph states and cluster states \cite{Li:2020,Istrati:2020}, Boson Sampling \cite{Lorado:2017}, Shor's algorithm \cite{Duan:2020}, a time-bit quantum relay \cite{Anderson:2020} and quantum key distribution \cite{Basset:2021,Chaiwongkhot:2020}.  

For the application of QDs to quantum communications systems, the maximum bit rate is limited by the radiative recombination time of electron-hole pairs in the QD.  While this bit rate may be increased by up to an order of magnitude through Purcell enhancement \cite{Kolchin:2015}, further improvements require the introduction of multiplexing schemes that allow the coupling of more than one channel into a single optical fiber.   The size variations that naturally occur in an ensemble of self-assembled semiconductor QDs lead to an inhomogeneous broadening of 10 to 30 meV \cite{Boggess:2001,Mohan:2010,Huffaker:2000}.  These fluctuations in the QD transition energy provide a built-in approach to wavelength division multiplexing in quantum networks, however the conventional method of triggering the emitter using a Rabi rotation is ineffective for a laser pulse detuned from the QD transition energy \cite{Ramachandran:2021}.  In such a case, multiplexing would come at the expense of considerable complexity as multiple, independently tunable laser sources would be required to trigger QDs emitting into different optical channels.  This would greatly impede the practicality and scalabilty of the resulting quantum photonic system.

Here we overcome this barrier by demonstrating the parallel inversion of QDs with an energy range spanning 15~meV using a single chirped laser pulse.  Our driving scheme uses adiabatic rapid passage (ARP), for which the exciton in the QD is inverted by evolving the optically-driven system through an anti-crossing in the dressed states.   ARP has been used to invert the exciton \cite{Simon:2011,Wu:2011,Mathew:2014,Kaldewey:2017,Ramachandran:2020} and biexciton \cite{Kaldewey2:2017,Kappe:2023} in single QDs as well as to trigger a single photon source \cite{Wei:2014}.  It has also recently been applied to the inversion of the biexciton in two distinct quantum dots \cite{Kappe:2023}, however the transition energies were only separated by 0.2~meV, in contrast to the large range reported here.  Our experiments are supported by theoretical simulations indicating that high-fidelity quantum state inversion is possible on a QD ensemble with realistic size variations using experimentally accessible values of laser pulse fluence and chirp.  Our findings verify the robustness of the ARP approach for implementation in multiplexed communication systems and will support the development of scalable quantum networks. 

\section*{Quantum Control using Chirped Laser Pulses}
 When a two-level quantum system is driven with an unchirped laser pulse (also referred to as a transform-limited pulse), the occupation of the upper level oscillates as a function of the pulse area ($\theta = \int_{-\infty}^{\infty}\frac{\mu E_p(t)}{\hbar} dt$).  This approach may be used to invert the quantum system, i.e. drive the occupation from the lower level to the upper level, provided the pulse area is exactly $\pi$. This process is referred to as a Rabi rotation.  The occupation of the upper state only reaches unity, corresponding to complete inversion of the quantum system, if the center frequency of the laser pulse is precisely resonant with the optical transition, a condition that is difficult to achieve for a system of QDs due to size variations within the ensemble.  In addition, the pulse area depends on the value of the dipole moment $\vec{\mu}$ of the QD, which also varies across the ensemble.  For these reasons, a Rabi rotation is only effective for quantum state inversion if the laser pulse is tailored to the optical properties of a single QD.  
 
\begin{figure}[ht]
\centering
\includegraphics[width=0.92\linewidth]{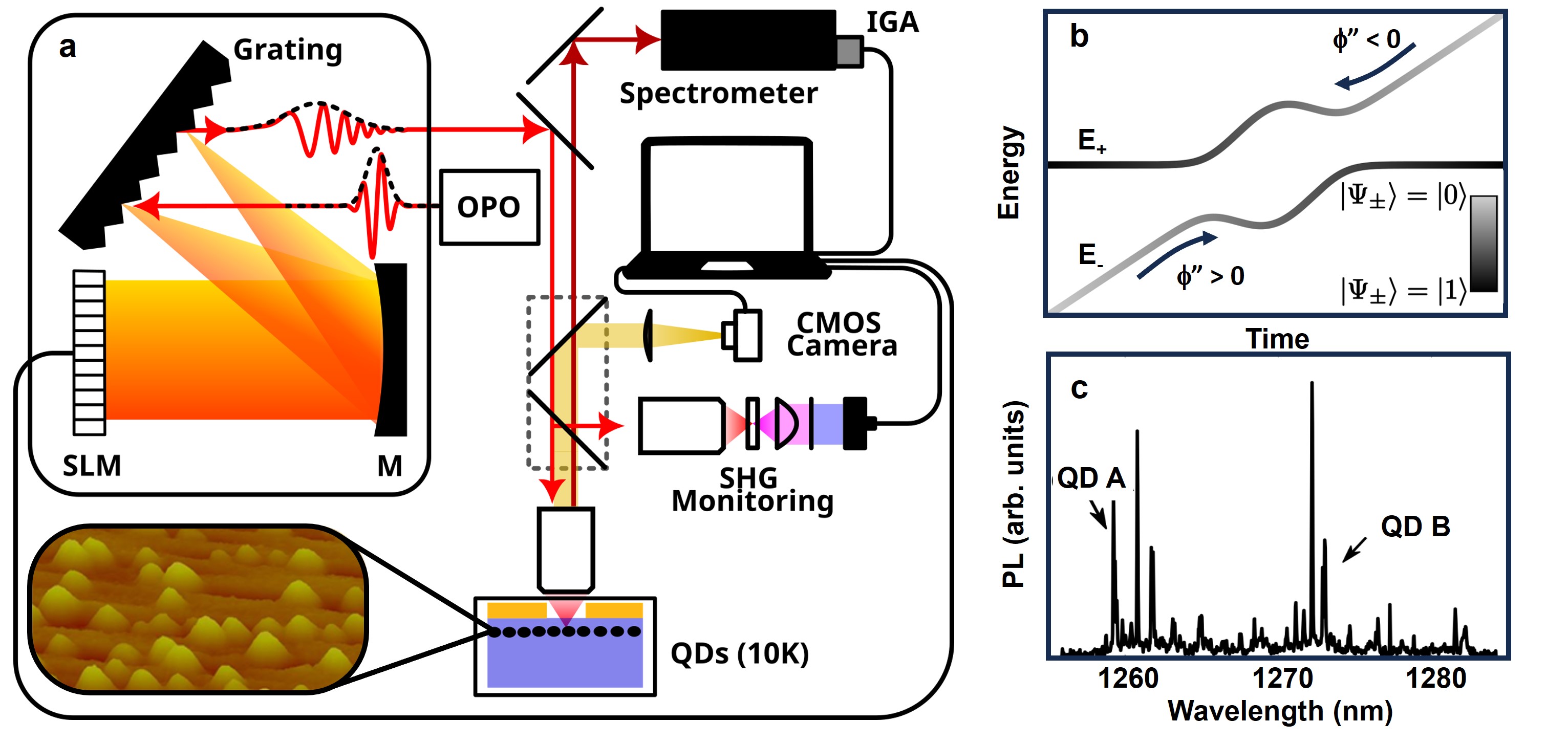}
\caption{{\bf a} Schematic diagram of the experimental setup.  Laser pulses from an optical parametric oscillator (OPO) are sent through a 4f pulse shaper containing a 128 pixel spatial light modulator (SLM) in the Fourier plane.   The pulse shaper is used to correct for phase distortions caused by optical components in the setup and to impose the chirp required for ARP.  Dispersion compensation is carried out using multiphoton intrapule interference phase scan (MIIPS) through measurement of the second harmonic generation (SHG) at an equivalent focus to the sample position. The GS emission from the QDs is detected using a 0.75~m spectrometer equipped with a liquid-N$_2$-cooled InGaAs array detector (IGA).  The inset shows an AFM image of the QDs, which have an average height of 9~nm and a lateral size of 20~nm with an areal density of 1$\times$10$^{10}$~cm$^{-2}$. The emission from approximately 30 QDs from the ensemble was isolated through the deposition of a metallic mask containing an array of apertures onto the sample surface. The QD sample is housed in a liquid-helium microscopy cryostat and at 10~K. {\bf b}   The temporal evolution of the energies (E$_{\pm}$) of the dressed states ($|\Psi_{\pm}>$) during ARP induced by a chirped laser pulse. {\bf c} PL from the GS transition of QDs in the interrogated aperture.  The emission peaks from QD~A and QD~B referred to in the text are indicated by arrows.}
\label{fig:Figure_1}
\end{figure}

 The parallel excitation of quantum emitters with distinct properties may instead be carried out using a frequency-swept laser pulse, which inverts the system using ARP.  In this case, the electric field of the control laser pulse is given by $E(t)=\frac{1}{2}E_p(t)\exp{[-i(\omega_l t+ \alpha t^2)]}$, where $E_p(t)$ is the pulse envelope, $\omega_l$ is the center frequency of the laser pulse and $\alpha$ is the rate of the frequency sweep.  In practice, this temporal frequency sweep is imposed in the frequency domain using a $4f$ pulse shaper (Fig.~\ref{fig:Figure_1}{\bf a}).  A grating is used to spatially separate the different frequency components and the time delay of each component is controlled using a spatial light modulator.  A linear time-varying frequency is obtained by introducing a spectral chirp $\phi ''$, where $\alpha=2\phi''/[\tau_0^4/(2\ln{(2)})^2+(2\phi'')^2]$, and $\tau_0$ is the duration of the associated transform-limited pulse.  The energies of the dressed states of a two-level quantum system driven by a chirped laser pulse are shown in Fig.~\ref{fig:Figure_1}{\bf b}.  The dressed states are denoted by $|\Psi_{\pm}\rangle$ (with energies E$_{\pm}$) and each consist of a time-dependent superposition of the unperturbed ground state ($|0\rangle$) and excited state ($|1\rangle$) of the exciton in the QD.  Within the adiabatic regime, the system remains in one of the dressed states throughout the laser pulse.  The quantum system inverts during the pulse because the dressed state itself reverses character, evolving from $|0\rangle$ to $|1\rangle$ or from $|1\rangle$ to $|0\rangle$ depending on the initial state of the exciton and the sign of the pulse chirp.  For our experiments, the QD starts in $|0\rangle$ and the pulse is positively chirped.  In this case, the system proceeds from left to right in Fig.~\ref{fig:Figure_1}{\bf b} on the lower-energy dressed state, leaving the exciton in state $|1\rangle$ at the end of the laser pulse.  

Unlike a Rabi rotation, ARP provides a robust approach to exciting excitons in inequivalent QDs because quantum state inversion is complete as long as the condition for adiabatic evolution is satisfied.  Under this condition, no transitions will occur between the dressed states during the optical pulse.  The adiabatic condition is given by $|\Delta \frac{d\Omega}{dt} - \Omega\frac{d\Delta}{dt}| \ll [\Omega^2 +\Delta^2]^{\frac{3}{2}}$ \cite{Shore:book}, where $\Omega(t) = \frac{\vec{\mu} \cdot \vec{E}(t)}{\hbar}$ is the Rabi frequency, $\Delta(t)$~=~$\Delta_0 - 2\alpha t$ is the instantaneous detuning and $\Delta_0$ the static detuning.  This condition may be satisfied for a range of detunings as long as the pulse area and chirp are sufficiently large.   As a result, in addition to insensitivity to pulse-to-pulse variations from the triggering laser source, ARP is tolerant to variations in the transition energies and dipole moments characteristic of self-assembled QDs. ARP was first demonstrated in atomic systems \cite{Melinger:1994,Vitanov:2001} and later in semiconductor QDs \cite{Simon:2011,Wu:2011}, including the achievement of subpicosecond control times for the inversion of excitons immune to phonon-mediated dephasing \cite{Mathew:2014,Ramachandran:2020}.  The robustness of ARP was recently verified experimentally for a single semiconductor QD \cite{Ramachandran:2021}, with high fidelity inversion being observed for a detuning of the laser pulse from the QD transition as large as 10 meV, spanning the transition energies in typical self-assembled QD ensembles.

\section*{Results and Discussion}

\begin{figure}[ht]
\centering
\includegraphics[width=0.85\linewidth]{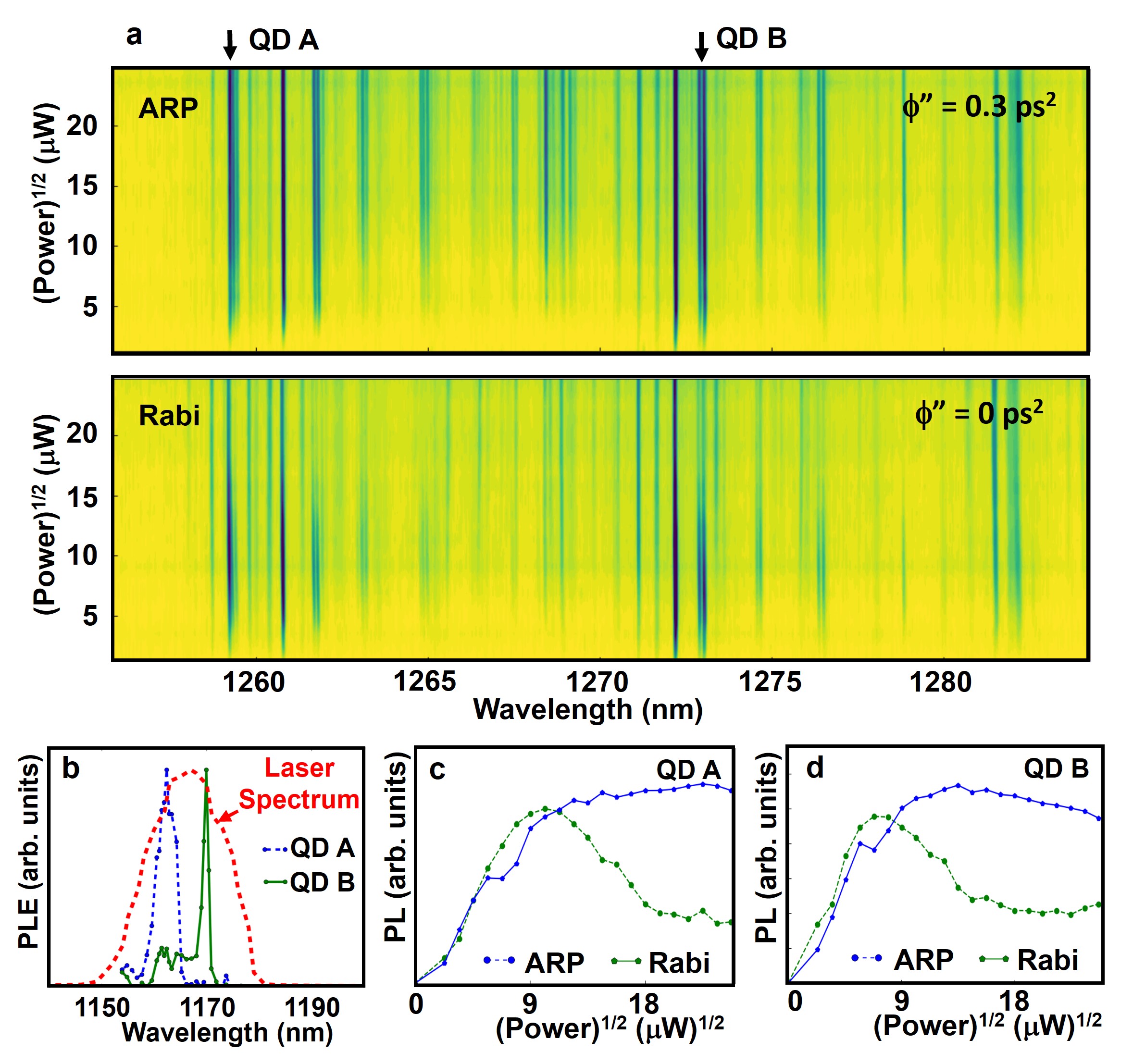}
\caption{{\bf a} PL intensity as a function of the square-root of the excitation power (proportional to pulse area) and GS emission wavelength for $\phi^{''}$~=~0.3~ps$^2$ (upper panel) and $\phi^{''}~=~0$ ps$^2$ (lower panel).  The color scale is identical in {\bf a} and {\bf b}.  Variation in the maximum intensity for different QDs reflects the differing optical coupling strengths into the detected emission mode as no optical cavity or solid-immersion lens is used to optimize light collection from the flat wafer.  For zero chirp, QDs undergo Rabi oscillations with varying contrast ratios for different QDs reflecting the sensitivity of quantum state inversion via Rabi rotations to detuning, while for $\phi^{''}$~=~0.3~ps$^2$ strong inversion characterized by saturation versus pulse area is observed for most QDs, consistent with robust state inversion via ARP. The GS emission peaks for QD~A and QD~B are indicated by arrows. {\bf b} Photoluminescence excitation spectrum for QD A and QD B, together with the laser spectrum. {\bf c}/({\bf d})  Same data as in {\bf a} for QD~A (QD~B) with ES transition wavelengths of 1162~nm (1171~nm) and GS emission wavelengths of 1259~(nm) (1273~nm). }
\label{fig:Figure_2}
\end{figure}

We report the demonstration of simultaneous ARP in an ensemble of InGaAs QDs with a ground-state (GS) optical transition centered at 0.976~eV (1270~nm), in the telecom  O-band at 10~K. In order to facilitate the spectral isolation of the emitted single photon stream from the scattered laser light, quantum control was carried out on the first excited state (ES) transition in the QDs, centered at 1.061~eV (1170~nm).  The ARP approach we use here could also be applied directly to the GS transition in conjunction with polarization-based filtering schemes \cite{Wei:2014}.  The inhomogeneous broadening of the GS transition is 30~meV, with a similar level expected for the ES \cite{Boggess:2001,Gamouras:2013}.  Approximately 30 QDs were isolated using a metallic mask with an array of apertures deposited onto the sample surface.  The results of quantum control experiments on these QDs are shown in Fig.~\ref{fig:Figure_2}{\bf a}.  The color scale indicates the strength of the GS emission as a function of wavelength and the square root of the driving laser power, which is proportional to the pulse area.   The different maximum PL intensities for different QD emission lines reflect variations in the location of the QDs within the aperture, which determines the efficiency of coupling of the optical emission into the objective lens and thus the efficiency of PL detection.   For these results, the laser spectrum was tuned to 1.063~eV (1166 nm), corresponding to the mean value of the ES transition energy for the ensemble.  For excitation with unchriped pulses, several of the QDs are observed to exhibit Rabi oscillations, however the contrast between oscillation maxima and minima varies considerably from QD to QD.  In contrast, for $\phi ''$~=~0.3~ps$^{2}$, the majority of QDs exhibit a saturation in the PL intensity versus the square root of the excitation power.  This observed saturation is a signature of ARP and robust state inversion.  

In order to further highlight the differing quantum state dynamics for optical driving with chirped and unchirped pulses, we selected two QDs within the ensemble (QD A and QD B, indicated by the arrows above the contour plots in Fig.~\ref{fig:Figure_2}{\bf a}).  The ES transitions for this pair of QDs are separated by 8~meV (occurring at 1162~nm and 1171~nm, respectively) and the spectrum of the control pulse was tuned in between these transitions, as shown in Fig.~\ref{fig:Figure_2}{\bf b}. For zero chirp, both QDs exhibit a damped Rabi oscillation. For chirped pulse excitation, the PL intensity saturates with increasing excitation power.  It is notable that the strength of the maximum PL emission for ARP is larger than that observed for Rabi oscillations for all values of pulse area.

\begin{figure}[ht]
\centering
\includegraphics[width=0.8\linewidth]{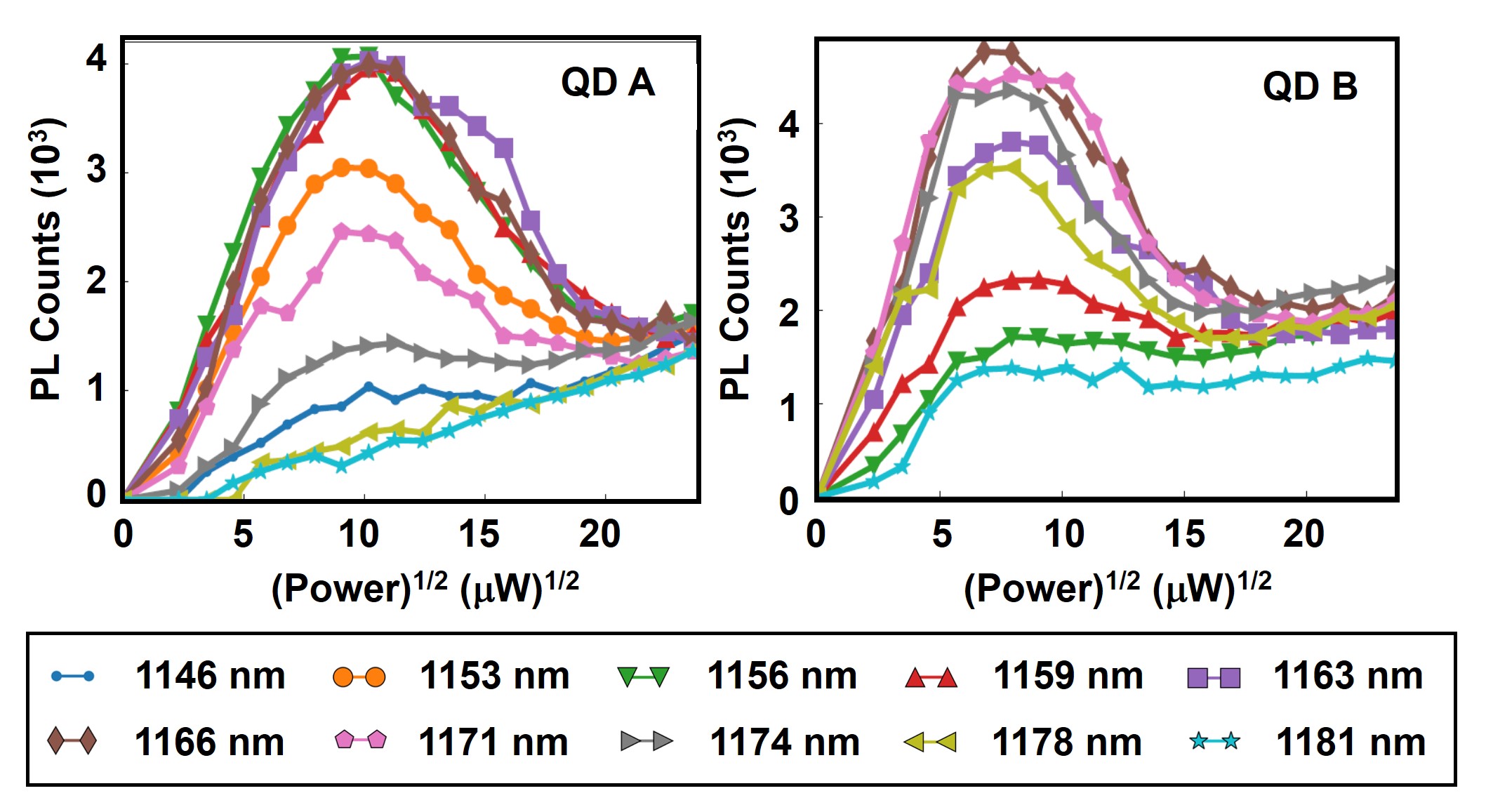}
\caption{PL intensity as a function of the square root of the excitation power for quantum control with $\phi^{''}~=~0$~ps$^2$ for different detunings of the center wavelength of the driving laser pulse from the ES transition in the QD (Left panel: QD~A; Right panel: QD B). The damped Rabi oscillation exhibits maximum contrast for resonant pumping of each QD, with progressively reduced contrast for increasing detuning.  A linearly-increasing background is caused by incoherent carrier pumping through nonlinear excitation channels such as two-photon absorption, which are negligible for ARP but contribute strongly for excitation with unchirped pulses due to the high laser peak intensity. 
 QD~A is observed to have a 25\% longer period of Rabi oscillation, indicating a lower dipole moment than QD~B. }
\label{fig:Figure_3}
\end{figure}

The poor contrast observed in the Rabi oscillations for many of the QDs in Fig.~\ref{fig:Figure_2}{\bf a} reflects the high sensitivity of quantum state inversion using this control process to variations in the optical properties of the QDs.  This sensitivity may be seen in Fig.~\ref{fig:Figure_3}, which shows the results of quantum control using unchirped pulses on QD~A and QD~B for a range of laser detunings.  As the pulse center photon energy deviates in either direction from the QD transition energy, the inversion of the exciton in the QD is rendered incomplete and the amplitude of the oscillation decreases, giving way to a linearly increasing background tied to incoherent filling of the QD states through nonlinear absorption channels.  This background is much stronger for excitation with a transform-limited pulse than for ARP due to the much larger peak intensities in the former case.  The poor contrast in the Rabi oscillations in Fig.~\ref{fig:Figure_2}{\bf c,d}, as well as the higher overall PL emission strength for control using ARP is expected as neither QD~A or QD~B is driven resonantly.  The Rabi periods indicate a 25\% difference in the dipole moment between the two QDs, yet ARP inverts the system effectively despite this difference.     
	
\begin{figure}[ht]
\centering
\includegraphics[width=0.9\linewidth]{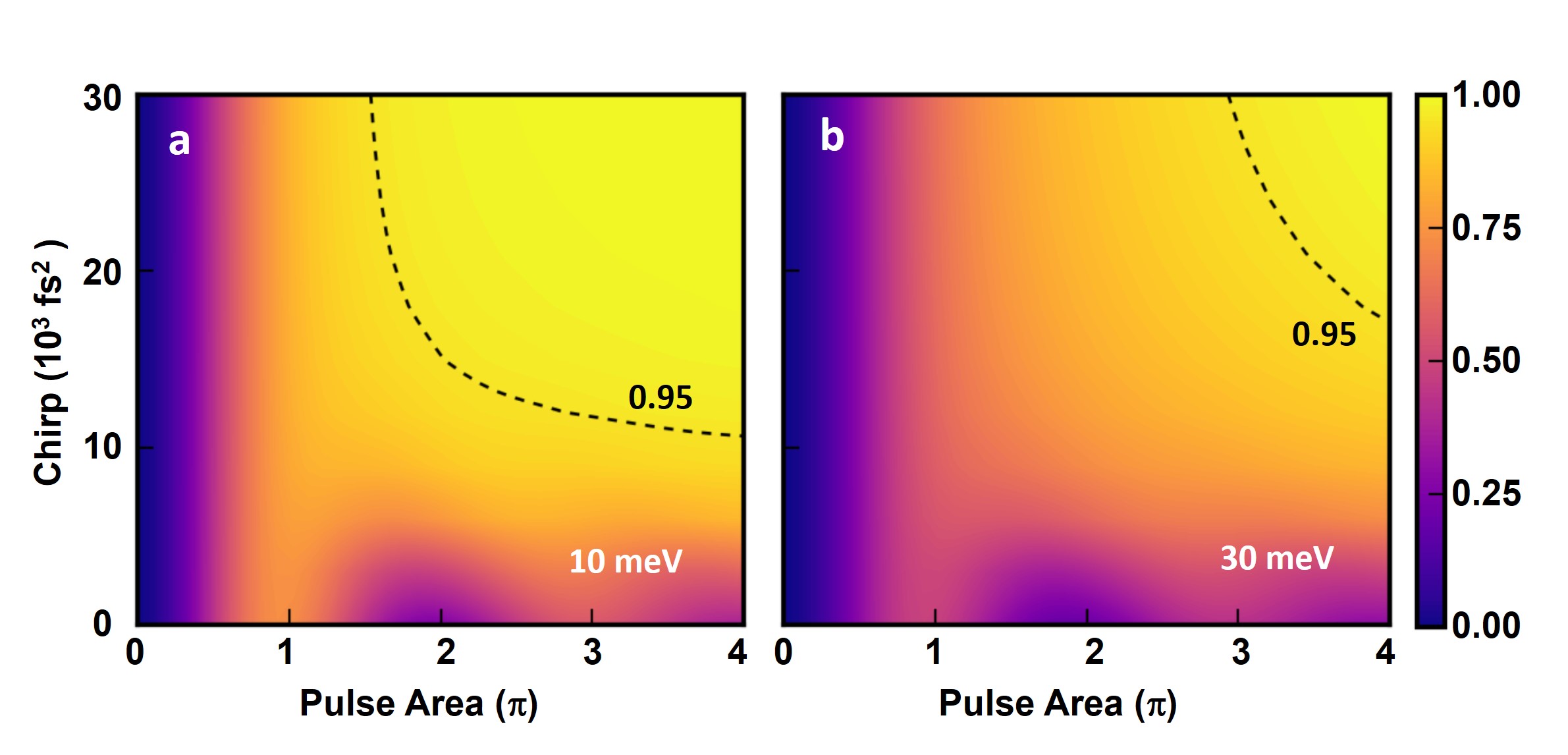}
\caption{Calculated total exciton occupation for an ensemble of 468 QDs as a function of chirp and pulse area for a Gaussian distribution of dipole moments (FWHM of 4 Debye, with a mean value of 25 Debye) and transition energies (taking the inhomogenous broadening as the FWHM) for two different inhomogeneous broadening levels: 10 meV ({\bf a}) and 30 meV ({\bf b}). The dotted curves indicate the conditions required to achieve a total occupation across the ensemble of 0.95.   The pulse area values on the x-axis correspond to the QDs that lie at the mean of the dipole moment distributions. }
\label{fig:Figure_4}
\end{figure}

To gain further insight into the level of robustness of the ARP control scheme, we carried out numerical simulations of the quantum state dynamics for an ensemble of 468 QDs. The results of these calculations are shown in Fig.~\ref{fig:Figure_4}.  The ensemble was taken to have a Gaussian distribution of transition energies with a full-width-at-half-maximum (FWHM) that was either 10~meV or 30~meV. The dipole moment distribution used had a FWHM of 4 Debye and a mean value of 25 Debye.  For zero pulse chirp, corresponding to calculated inversion values along the x-axis in Fig.~\ref{fig:Figure_4}, a damped Rabi oscillation is observed for which the maximum inversion is 0.72 in Fig.~\ref{fig:Figure_4}{\bf a} and 0.47 in Fig.~\ref{fig:Figure_4}{\bf b}.  The low value of the maximum exciton occupation and the low contrast of the Rabi oscillations for zero chirp reflect the high sensitivity of quantum state inversion via a Rabi rotation to variations in the QD optical properties, an effect that becomes worse the larger the degree of QD inhomogeneity. As the laser chirp is increased from zero, the exciton occupation increases.  The dashed curves in Fig.~\ref{fig:Figure_4} indicate a calculated total exciton occupation of 0.95, showing that a high exciton inversion is accessible for sufficiently large values of the pulse chirp and pulse area.  An occupation of 0.99 occurs above a threshold pulse area of 2.45$\pi$ (3.74$\pi$) and chirp of 0.018~ps$^2$ (0.026~ps$^2$) for an inhomogeneous broadening of 10~meV (30~meV).  While similar results were obtained in earlier numerical simulations of optically-controlled QD ensembles \cite{Schmidgall:2010}, a much narrower distribution of transition energies was used (2 meV), in contrast to the realistic QD size distribution considered here \cite{Boggess:2001}.          

The ability to simultaneously trigger QDs over a wide range of transition energies, as we demonstrate here, would greatly facilitate the development of high-bandwidth quantum networks.  One could assign distinct quantum optical channels across the inhomogeneous bandwidth using narrow bandpass filters to extract the single photon emission from selected QDs. These QDs could be isolated in different cavities and pumped using the sample driving laser or one could exploit a dense ensemble integrated into a single broadband cavity. 
A channel separation of 0.75~meV would optimally exploit the resolution of standard 4f pulse shaping systems, which could be implemented in conjunction with spectral engineering techniques \cite{Wilbur:2022,He:2019,Koong:2021,Bracht:2021}.  Such an approach would increase the bandwidth of quantum optical communication systems by 20-fold while simultaneously exploiting highly-developed classical communication hardware for wavelength division multiplexing, increasing the scalability of quantum networks without adding significant complexity.

\section*{Conclusions}

In this work, we demonstrate the simultaneous inversion of an ensemble of $>$10 QDs with a wide range of optical properties using a single chirped laser pulse.  
While some progress in the parallel driving of distinct QDs has been made in recent years, including the application of general pulse shape engineering to arbitrary qubit rotations \cite{Mathew:2011,Gamouras:2012,Gamouras:2013,Mathew:2015,Kappe:2023}, these experiments have been limited to a pair of quantum dots, in contrast to the work reported here.  The implementation of ARP on distinct emitters could be applied to parallel quantum state initialization in QD-based qubits and to laser triggering of QD sources of single photons, adding to the toolkit for coherent optical control in solid-state systems.  The insensitivity of ARP to the optical properties of individual QDs was highlighted with measurements on two QDs with transition energies separated by 8~meV and with a 25\% difference in dipole moment.  Theoretical simulations were also performed, indicating that inversion of an ensemble with an inhomogeneous bandwidth as large as 30~meV would be possible using experimentally accessible values of chirp and pulse area. We estimate that the robustness of ARP would enable the implementation of up to 20 separate communication channels, supporting the development of high-bandwidth, multiplexed quantum networks.

\section*{Methods}

\subsection*{Quantum Dot Sample}
The In(Ga)As/GaAs QDs studied in this work were grown using molecular beam epitaxy under conditions optimized to achieve a GS transition in the O-band at cryogenic temperature.  The QD layer was deposited under In-rich conditions on top of a 200~nm GaAs buffer, followed by overgrowth with In$_{0.2}$Ga$_{0.8}$As and a 65~nm GaAs capping layer.  The single monolayer of QDs was surrounded by AlGaAs carrier blocking layers.  The areal density of the QDs is ~1$\times$10$^{10}$~cm$^{-2}$. 
 The optical transitions in the QDs were characterized using PL and PL excitation measurements.  The GS PL emission peak for the QD ensemble was centered at 1270~nm with the ES transition centered at 1166~nm at 10~K. A metallic mask with an array of apertures was deposited onto the sample surface and used to isolate a set of QDs.  The experiments reported here were carried out under excitation through a 0.6$\mu$m aperture, leading to the simultaneous excitation of approximately 30 QDs. 

\subsection*{Quantum Control Experiments}
 The broadband laser pulses for optical control were generated using an optical parametric oscillator with a repetition rate of 76 MHz, tunable from 1100~nm to 1600~nm.  Correction for phase distortions caused by optical components in the setup was carried out using a 4f pulse shaper equipped with multi-photon intrapulse interference phase scan, resulting in a duration of $\tau_0$~=~120~fs for the unchirped pulse with a time-bandwidth product within 5\% of the transform limit.   For quantum control using ARP, the same $4f$ pulse shaper was used to impose a spectral chirp of  $\phi ''$~=~0.3~ps$^{2}$.  Quantum control was carried out on the ES transition in the QDs, with quantum state readout through detection of the PL from the GS transition following non-radiative relaxation of excited carriers, providing a proxy for the ES occupation at the end of the laser pulse.   The laser pulses were circularly polarized to suppress biexciton dynamics, and the large energy separation between the GS and ES transitions (85 meV) and between the first excited state transition and the wetting layer (310 meV) aided in avoiding unintended excitations.  A high numerical aperture microscope objective (NA 0.7, 100x) was used to focus the excitation beam onto the sample and to collect the emitted PL. The GS PL emission from the QDs was spectrally resolved and detected using a 0.75~m monochromator with a resolution of 30~$\mu$eV and a liquid-N$_{2}$-cooled InGaAs array detector. For all experiments, the sample was held at 10 K in a continuous flow helium cryostat.

\bibliography{sample}

\section*{Acknowledgements}

This research was supported by the Natural Sciences and Engineering Research Council of Canada, Grant No. RGPIN-2020-06322, and the National Research Council of Canada Internet of Things: Quantum Sensors Challenge Programs QSP 042 and QSP 083.

\section*{Author contributions statement}

K.C.H. conceived the experiment(s), A.R. carried out the experiments with assistance from A.M., A.R. and G.W. analyzed the experimental data, S.O. and D.G.D. carried out the growth of the QD structure studied, K.C.H., A.R. and G.W. wrote the manuscript with input from all authors. 

\section*{Data Availability}
All data generated or analysed during this study are included in this published article. 

\section*{Additional information}

The authors declare no competing interests.

%\begin{figure}[ht]
%\centering
%\includegraphics[width=\linewidth]{stream}
%\caption{Legend (350 words max). Example legend text.}
%\label{fig:stream}
%\end{figure}

\end{document}